\documentclass[11pt,superscriptaddress,aps,prd,preprint]{revtex4}
\usepackage{graphicx}
\usepackage{amsmath}
\usepackage{slashed}	
\usepackage[latin1]{inputenc}

\begin{document}
\title{Radiatively induced CPT-odd Chern-Simons term in massless QED}
\author{J. F. Assun\c c\~ao}
\author{T. Mariz}
\affiliation{Instituto de F\'\i sica, Universidade Federal de Alagoas, 57072-900, Macei\'o, Alagoas, Brazil}
\email{ jfassuncao,tmariz@fis.ufal.br}
\date{\today}
\begin{abstract}
In this work, we study the radiative generation of the CPT-odd Lorentz-violating Chern-Simons term, arising from massless fermions. For this, we calculate the vacuum polarization tensor using the 't Hooft-Veltman regularization scheme, in which the result obtained for the coefficient of the Chern-Simons term is $(k_{AF})_\mu=-\frac{e^2}{4\pi^2}\,b_\mu$. This result leads us precisely to the same conductivity found in Weyl semimetals, i.e., the 't Hooft-Veltman regularization scheme is the correct one to be used in this context. We also discuss the temperature dependence of $(k_{AF})_\mu$, in which at high temperature, $(k_{AF})_0\to0$ and $(k_{AF})_i\to-\frac{e^2}{4\pi^2}b_i$. In the context of Weyl semimetals, these results are in accordance with the fact that the chiral magnetic current $j^\alpha=(k_{AF})_0\epsilon^{0\alpha\lambda\rho}\partial_\lambda A_\rho$ vanishes at high temperature, whereas the anomalous Hall current $j^\alpha=(k_{AF})_i\epsilon^{i\alpha\lambda\rho}\partial_\lambda A_\rho$ remains unaffected by the finite temperature.
\end{abstract}
\maketitle

\section{Introduction}

Studies of CPT and Lorentz violation in the standard model \cite{Colladay:1996iz,Colladay:1998fq} have been extensively considered in the literature in the last years \cite{Kostelecky:2008ts}. The effective field theory used for these investigations is the standard-model extension, where the corresponding CPT-odd Lorentz-violating quantum electrodynamics \cite{Kostelecky:2001jc} is given by the Lagrangian
\begin{eqnarray}\label{lvqed}
\mathcal{L}&=&-\textstyle{\frac14}F_{\mu\nu}F^{\mu\nu}+\textstyle{\frac12}(k_{AF})_\mu\epsilon^{\mu\nu\lambda\rho}A_\nu F_{\lambda\rho} \nonumber\\
&&+\bar{\psi}(i\slashed{D}+\textstyle{\frac12}g^{\lambda\rho\mu}\sigma_{\lambda\rho}D_\mu-m-\slashed{b}\gamma_5)\psi,
\end{eqnarray}
where $F_{\mu\nu}=\partial_\mu A_\nu-\partial_\nu A_\mu$ and $D_\mu=\partial_\mu+ieA_\mu$. We have omitted the CPT-odd terms that are governed by the coefficients $e_\mu$, $f_\mu$, and $a_\mu$, which can be removed from the Lagrangian by using an appropriate redefinition of the spinor \cite{Colladay:2002eh}, as well as the CPT-even terms contracted with the coefficients $(k_F)_{\mu\nu\lambda\rho}$, $c_{\mu\nu}$, $d_{\mu\nu}$, and $H_{\mu\nu}$.

It is worth mentioning that the term $\bar\psi\slashed{b}\gamma_5\psi$, the last term of Eq.~(\ref{lvqed}), is the one responsible for the radiative generation of the Chen-Simons term \cite{Jackiw:1999yp}, the second term of Eq.~(\ref{lvqed}), so that $(k_{AF})_\mu \propto b_\mu$. In this way, we can write $(k_{AF})_\mu=C\,b_\mu$, where $C$ is a proportionality constant to be determined. Thus, these terms must have identical C, P, and T transformation properties, as listed in Table 1. Note that the term governed by the coefficient $g_{\lambda\rho\mu}$, the fourth term of Eq.~(\ref{lvqed}), has also the same discrete transformations. In fact, this term generates a higher-derivative Chern-Simons term in which $(k_{AF})_\mu \to {(k_{AF})_\mu}^{\alpha\beta}\partial_\alpha\partial_\beta\propto\epsilon_{\mu\nu\lambda\rho}{\cal G}^{\nu\lambda\rho\alpha\beta}\partial_\alpha\partial_\beta$, where ${\cal G}^{\nu\lambda\rho\alpha\beta}= g^{\nu\lambda\alpha}g^{\beta\rho}+g^{\lambda\rho\alpha}g^{\beta\nu}+g^{\rho\nu\alpha}g^{\beta\lambda}$. For more details, see Ref.~\cite{Mariz:2010fm}.

\medskip
\centerline{\begin{tabular}{||c|c|c|c||c||}
\hline
& C & P & T & CPT \\ \hline 
$b_j,g_{j0l},g_{jk0},(k_{AF})_j$ & $+$ & $+$ & $-$ & $-$ \\ \hline
$b_0,g_{j00},g_{jkl},(k_{AF})_0$ & $+$ & $-$ & $+$ & $-$ \\ \hline
\end{tabular}}
\medskip
\centerline{\small Table 1: Discrete symmetry properties.}
\smallskip 

The aim of the present work is to investigate the radiative generation of the CPT-odd Lorentz-violating Chern-Simons term arising from massless fermions. This study was already addressed in the literature \cite{PerezVictoria:1999uh,Brito:2008ec}, where it has been argued in \cite{Brito:2008ec} that the result $C_1=-\frac{e^2}{16\pi^2}$ is finite and determined, at least in three types of regularization schemes: cutoff, dimensional, and temperature regularization. However, recently, a new result has been found, namely, $C_2=-\frac{e^2}{4\pi^2}$, which was calculated by using the path integral method (Fujikawa's method) \cite{Zyuzin:2012tv} and the Kubo formula \cite{Goswami:2012db}, in the context of condensed-matter physics. It is worth commenting the following question raised in Ref.~\cite{Goswami:2012db}: why do the available calculations of the polarization tensor \cite{Jackiw:1999yp,PerezVictoria:1999uh,Brito:2008ec} fail to obtain the correct coefficient $C_2$? We will see that when we use the 't Hooft-Veltman (dimensional) regularization scheme \cite{'tHooft:1972fi}, the result $C_2$ is readily obtained. 

The motivation for these recent works \cite{Zyuzin:2012tv,Goswami:2012db} stems from the fact that certain materials known as Weyl semimetals~\cite{Burkov} are perfectly described by the Lorentz-violating fermionic Hamiltonian \cite{Grushin:2012mt}
\begin{equation}\label{fH}
{\cal H}_\psi = \bar{\psi}(-i\partial_i\gamma^i+\slashed{b}\gamma_5)\psi,
\end{equation}
i.e. by the Lorentz-violating fermionic Lagrangian
\begin{equation}\label{fL}
{\cal L}_\psi = \bar\psi(i\slashed{\partial}-\slashed{b}\gamma_5)\psi,
\end{equation}
and consequently by the radiatively induced Lorentz-violating Chern-Simons Lagrangian. For example, the conductivity found in the Weyl semimetal proposed in~\cite{Burkov}, which is based on a periodic array of alternating layers of topological insulators\footnote{For a review on topological insulators, we refer the reader to Ref.~\cite{Hasan:2010xy}.} and ordinary insulators, is the same we can obtain from the Chern-Simons action, with the result $C_2$ (see also discussions in~\cite{Zyuzin:2012vn}). In order to observe this, we first calculate the current density,
\begin{equation}
j^\alpha=\frac{\delta S_{CS}}{\delta A_\alpha}=-\frac{e^2}{2\pi^2}b_\mu\epsilon^{\mu\alpha\lambda\rho}\partial_\lambda A_\rho,
\end{equation}
in which, for a pure spacelike coefficient $b^\mu=(0,0,0,b_z)$, we can write $j^1=\frac{e^2}{2\pi^2}b^3\epsilon^{3102}E^2$ (i.e., $j_x=\sigma_{xy}E_y$), so that, finally for the conductivity, we have 
\begin{equation}\label{cond}
\sigma_{xy}=\frac{e^2}{2\pi^2}b_z.
\end{equation}
Note that with $C_1$ the resulting conductivity is four times smaller than the expected answer (\ref{cond}), as pointed out in Ref.~\cite{Goswami:2012db}. 

The structure of the paper is as follows. In the next section, we calculate the vacuum polarization tensor, and consequently the generation of the Chern-Simons term, by using the 't Hooft-Veltman regularization scheme. For this, we will take into account nonperturbative approach in the coefficient $b_\mu$. In the third section, we discuss the temperature dependence of the coefficient $(k_{AF})_\mu$. We will see that, with $(k_{AF})_\mu=C\,b_\mu$, at high temperature, $(k_{AF})_0\to0$ and $(k_{AF})_i\to-\frac{e^2}{4\pi^2}b_i$, i.e. the parity symmetry is restored. In the context of Weyl semimetals, this confirms the fact that the chiral magnetic current $j^\alpha=(k_{AF})_0\epsilon^{0\alpha\lambda\rho}\partial_\lambda A_\rho$ vanishes at high temperature \cite{Zyuzin:2012vn}, whereas the anomalous Hall current $j^\alpha=(k_{AF})_i\epsilon^{i\alpha\lambda\rho}\partial_\lambda A_\rho$ remains unaffected by the temperature \cite{Goswami:2012db}.

\section{Induced Chern-Simons term}\label{CS}

In this section we are interested in studying the generation of the Lorentz-violating Chern-Simons term arising from massless fermions. For this, let us consider the fermionic Lagrangian (\ref{fL}), with the introduction of the electromagnetic field $A_\mu$ (by performing the replacement  $\partial_\mu\to D_\mu$), rewritten as
\begin{equation}\label{fLp}
{\cal L}'_\psi = \bar\psi(i\slashed{\partial}-\slashed{b}\gamma_5-e\slashed{A})\psi.
\end{equation}
Thus, the corresponding generating functional is
\begin{equation}
Z[A_\mu] = \int D\bar\psi D\psi e^{i\int d^4x{\cal L}'_\psi} = e^{iS_\mathrm{eff}},
\end{equation}
so that, by integrating over the fermions, we obtain the one-loop effective action 
\begin{equation}\label{Seff}
S_\mathrm{eff} = -i\mathrm{Tr}\ln(\slashed{p}-\slashed{b}\gamma_5-e\slashed{A}).
\end{equation}
Here, $\mathrm{Tr}$ stands for the trace over the Dirac matrices, as well as the trace over the integration in momentum and coordinate spaces.

In order to single out the quadratic terms in $A_\mu$ of the effective action,  we initially rewrite the expression (\ref{Seff}) as
\begin{equation}
S_\mathrm{eff}=S_\mathrm{eff}^{(0)}+\sum_{n=1}^\infty S_\mathrm{eff}^{(n)},
\end{equation}
where $S_\mathrm{eff}^{(0)}=-i\mathrm{Tr}\ln(\slashed{p}-\slashed{b}\gamma_5)$ and 
\begin{equation}
S_\mathrm{eff}^{(n)} = \frac{i}{n}\mathrm{Tr}\left(\frac{1}{\slashed{p}-\slashed{b}\gamma_5}e\slashed{A}\right)^n.
\end{equation}
Then, after evaluating the trace over the coordinate space, by using the commutation relation $A_\mu(x) G_b(p)=G_b(p-i\partial)A_\mu(x)$ and the completeness relation of the momentum space, for the quadratic action $S_\mathrm{eff}^{(2)}$, we have
\begin{equation}\label{Seff2}
S_\mathrm{eff}^{(2)} = \frac{1}{2}\int d^4x\, \Pi^{\mu\nu} A_\mu A_\nu,
\end{equation}
where
\begin{equation}\label{Pi}
\Pi^{\mu\nu} = ie^2\int\frac{d^{4}p}{(2\pi)^4}\mathrm{tr}\,G_{b}(p)\gamma^\mu G_{b}(p-i\partial)\gamma^\nu,
\end{equation}
with
\begin{equation}\label{Gb}
G_b(p) = \frac{1}{\slashed{p}-\slashed{b}\gamma_5}
\end{equation}
being the Feynman propagator. Note that the derivative contained in $\Pi^{\mu\nu}$ acts only on the first gauge field $A_\mu$, in Eq.~(\ref{Seff2}).

Our next step is to calculate the vacuum polarization tensor (\ref{Pi}), using the 't Hooft-Veltman regularization scheme. For this, we first extend the $4$-dimensional spacetime to a $D$-dimensional one, so that $d^4p/(2\pi)^4$ goes to $\mu^{4-D}d^D\bar p/(2\pi)^D$, where $\mu$ is an arbitrary parameter that identifies the mass scale. In the following, given the anticommutation relation $\{\bar\gamma^\mu,\bar\gamma^\nu\}=2\bar g^{\mu\nu}$, with the contraction $\bar g_{\mu\nu}\bar g^{\mu\nu}=D$, we split the $D$-dimensional Dirac matrices $\bar\gamma^\mu$ and the $D$-dimensional metric tensor $\bar g^{\mu\nu}$ into $4$-dimensional parts and $(D-4)$-dimensional parts, i.e., $\bar\gamma_\mu=\gamma_\mu+\hat\gamma_\mu$ and $\bar g^{\mu\nu}=g^{\mu\nu}+\hat g^{\mu\nu}$, such that now the Dirac matrices satisfy the anticommutation relations 
\begin{subequations}\label{gamma}
\begin{equation}
\{\gamma^\mu,\gamma^\nu\}=2g^{\mu\nu},\hspace{0.4cm} \{\hat\gamma^\mu,\hat\gamma^\nu\}=2\hat g^{\mu\nu},\hspace{0.4cm} \{\gamma^\mu,\hat\gamma^\nu\}=0,
\end{equation}
and consequently the metric tensors have the contractions $g_{\mu\nu}g^{\mu\nu}=4$, $\hat g_{\mu\nu}\hat g^{\mu\nu}=D-4$, and $g_{\mu\nu}\hat g^{\mu\nu}=0$. Indeed, the most significant change found in this regularization is the introduction of the commutation relation 
\begin{equation}
[\gamma_5,\hat\gamma^\mu]=0
\end{equation}
and the maintenance of the anticommutation relation 
\begin{equation}
\{\gamma_5,\gamma^\mu\}=0.
\end{equation}
\end{subequations}
For further information about this issue, see Sec.~IV of Ref.~\cite{Buras:1998raa}. 

Taking into account the above considerations, let us rationalize the propagator~(\ref{Gb}), as follows: 
\begin{equation}\label{Gb2}
G_{b}(p) = \frac{1}{\bar{\slashed{p}}-\slashed{b}\gamma_5}\;\frac{\bar{\slashed{p}}-\slashed{b}\gamma_5}{\bar{\slashed{p}}-\slashed{b}\gamma_5}\;\frac{\bar{\slashed{p}}+\slashed{b}\gamma_5}{\bar{\slashed{p}}+\slashed{b}\gamma_5}\;\frac{\bar{\slashed{p}}+\slashed{b}\gamma_5}{\bar{\slashed{p}}+\slashed{b}\gamma_5},
\end{equation}
where we have promoted $\slashed{p}$ to a quantity in $D$ dimensions, $\slashed{p}\to\bar{\slashed{p}}=\bar p_\mu \bar\gamma^\mu$, whereas $\slashed{b}$ is maintained in $4$ dimensions. In fact, all quantities of the external particles, such as $\slashed{A}$ and $\slashed{\partial}$, are maintained in 4 dimensions. Thus, the evaluation of Eq.~(\ref{Gb2}), by using the anticommutation and commutation relations (\ref{gamma}) as well as the separation $\bar p^\mu=p^\mu+\hat p^\mu$ (with $\hat p_\mu\gamma^\mu=0=p_\mu\hat\gamma^\mu$), takes the form
\begin{eqnarray}\label{Gb3}
G_b(p) = \frac{\bar p^2+b^2+2(p \cdot b)\gamma_5+[\hat{\slashed{p}},\slashed{b}]\gamma_5}{(\bar p^2-b^2)^2-4[(p\cdot b)^2-p^2b^2]}(\bar{\slashed{p}}+\slashed{b}\gamma_5).
\end{eqnarray}
Note that we can rewrite the above expression as
\begin{equation}\label{Gb4}
G_b(p) = \frac{\bar p^2+b^2+2(\bar p \cdot b)\gamma_5+[\hat{\slashed{p}},\slashed{b}]\gamma_5}{(\bar p-b)^2(\bar p+b)^2-4\hat p^2b^2}(\bar{\slashed{p}}+\slashed{b}\gamma_5),
\end{equation}
where we have consider $p\cdot b=\bar p\cdot b$ and $p^2=\bar p^2-\hat p^2$ (i.e., $\hat p\cdot b=0=\hat p\cdot p$). However, it is more convenient to consider the expansion of the Eq.~(\ref{Gb4}) in terms of $4\hat p^2b^2$, so that we write
\begin{equation}\label{Bg5}
G_b(p) = S_b(p)+\frac{4\hat p^2b^2}{(\bar p-b)^2(\bar p+b)^2}S_b(p) + \cdots,
\end{equation}
where $S_b(p)=S_{b1}(p)+S_{b2}(p)$, with
\begin{subequations}\label{Sbpartes}
\begin{eqnarray}
\label{Sb1}S_{b1}(p) &=& \frac{\bar p^2+b^2+2(\bar p \cdot b)\gamma_5}{(\bar p-b)^2(\bar p+b)^2}(\bar{\slashed{p}}+\slashed{b}\gamma_5),\\
\label{Sb2}S_{b2}(p) &=& \frac{[\hat{\slashed{p}},\slashed{b}]\gamma_5}{(\bar p-b)^2(\bar p+b)^2}(\bar{\slashed{p}}+\slashed{b}\gamma_5).
\end{eqnarray}
\end{subequations}
It is not difficult to see that the second and other higher order terms of the propagator~(\ref{Bg5}), by power counting, contribute with finite terms to the polarization tensor (\ref{Pi}). Therefore, they will be vanish when it happens the contraction $\bar g_{\mu\nu}\hat g^{\mu\nu}=D-4$, in the limit of~$D\to4$. 

Thus, by considering $i\partial\to k$, Eq.~(\ref{Pi}) can be rewritten only in terms of the propagators (\ref{Sbpartes}), i.e., $\Pi^{\mu\nu}=\Pi_{11}^{\mu\nu}+\Pi_{12}^{\mu\nu}+\Pi_{21}^{\mu\nu}+\Pi_{22}^{\mu\nu}$, with
\begin{equation}
\Pi_{ij}^{\mu\nu} = ie^2\mu^{4-D}\int\frac{d^D\bar p}{(2\pi)^D}\mathrm{tr}\,S_{bi}(p)\gamma^\mu S_{bj}(p-k)\gamma^\nu.
\end{equation}
In order to perform the above integrations, we first combine the denominators by employing the Feynman parameters. Then, for the first contribution, $\Pi_{11}^{\mu\nu}$, we have
\begin{eqnarray}\label{Pi11}
\Pi_{11}^{\mu\nu} &=& \int_0^1dx\int_0^{1-x}dy\int_0^{1-x-y}dz\;\mu^{4-D}\int \frac{d^D\bar p}{(2\pi)^D}\nonumber\\
&&\times\frac{6ie^2}{(\bar p^2-M^2)^4}\mathrm{tr}\{[\bar q^2+ b^2+2(\bar q\cdot b)\gamma_5](\bar{\slashed{q}}+\slashed{b}\gamma_5) \nonumber\\
&&\times\bar \gamma^\mu[\bar q_1^2+ b^2+2(\bar q_1\cdot b)\gamma_5](\bar{\slashed{q}}_1+\slashed{b}\gamma_5)\bar\gamma^\nu\},
\end{eqnarray}
where
\begin{eqnarray}
M^2 &=& -4(1-x-z)(x+z) b^2+4[(1-x-y)x \nonumber\\
&&-(x+y)z] b\cdot k-(1-x-y)(x+y) k^2,
\end{eqnarray}
$\bar q^\mu=\bar p^\mu+t^\mu$, and $\bar q_1^\mu=\bar p^\mu+t_1^\mu$, with 
\begin{eqnarray}
t^\mu &=& -b^\mu(1-2x-2y)+k^\mu(1-x-y)\\ 
t_1^\mu &=& -b^\mu(1-2x-2y)+k^\mu(1-x-y)-k^\mu.
\end{eqnarray}

Let us focus on the odd part of Eq.~(\ref{Pi11}), which is the one responsible for the generation of the Chern-Simons term, by writing $\Pi_{11}^{\mu\nu}=\Pi_{11|even}^{\mu\nu}+\Pi_{11|odd}^{\mu\nu}$. In this way, after the calculation of the trace, using Eqs.~(\ref{gamma}), we obtain
\begin{equation}\label{Pi11odd}
\Pi_{11|odd}^{\mu\nu} = 24e^2\int_0^1dx\int_0^{1-x}dy\int_0^{1-x-y}dz(I^{\mu\nu}_1+I^{\mu\nu}_2),
\end{equation}
with
\begin{eqnarray}
I^{\mu\nu}_1 &=& 2\mu^{4-D}\int \frac{d^D\bar p}{(2\pi)^D} \frac{\epsilon^{\kappa\lambda\mu\nu}k_\kappa \bar q_\lambda}{(\bar p^2-M^2)^4} \\
&&\times[b^2(b\cdot \bar q)+ b^2(b\cdot \bar q_1)+(b\cdot \bar q_1)q^2+(b\cdot \bar q)\bar q_1^2] \nonumber
\end{eqnarray}
and
\begin{eqnarray}
I^{\mu\nu}_2 &=& \mu^{4-D}\int \frac{d^D\bar p}{(2\pi)^D} \frac{\epsilon^{\kappa\lambda\mu\nu}b_\kappa k_\lambda}{(\bar p^2-M^2)^4} \\
&&\times[4(b\cdot \bar q)(b\cdot \bar q_1)+(b^2+\bar q^2)(b^2+\bar q_1^2)], \nonumber
\end{eqnarray}
where we have considered a Hermitian $\gamma_5$, so that $\mathrm{tr}\,\gamma^\kappa\gamma^\lambda\gamma^\mu\gamma^\nu=4i\epsilon^{\kappa\lambda\mu\nu}$. Now, after we integrate over the momentum $\bar p$, expand the result around $D=4$, and consider $k^2=0=b\cdot k$, $\Pi_{11|odd}^{\mu\nu}\to\Pi_{11|CS}^{\mu\nu}$, i.e., we obtain
\begin{eqnarray}\label{Pi11CS}
\Pi_{11|CS}^{\mu\nu} &=& -\frac{ie^2}{4\pi^2}\epsilon^{\kappa\lambda\mu\nu}b_\kappa k_\lambda\int_0^1dx\int_0^{1-x}dy\int_0^{1-x-y}dz \nonumber\\ 
&&\times\frac{1-6z+6[z^2+2xz-(1-x)x]}{(1-x-z)(x+z)}.
\end{eqnarray}
Finally, when the integrals over the Feynman parameters are evaluated, we have 
\begin{equation}\label{Pi11CS}
\Pi_{11|CS}^{\mu\nu}=0.
\end{equation}

In the following, let us take into account the contributions $\Pi_{12}^{\mu\nu}$ and $\Pi_{21}^{\mu\nu}$. Note that, given the expression
\begin{equation}
\Pi_{12}^{\mu\nu}(k) = ie^2\mu^{4-D}\int \frac{d^D\bar p}{(2\pi)^D}\mathrm{tr}\,S_{b1}(p)\gamma^\mu S_{b2}(p-k)\gamma^\nu,
\end{equation}
we can easily observe that $\Pi_{21}^{\nu\mu}(k)=\Pi_{12}^{\nu\mu}(-k)$. Therefore, we simply calculate, e.g., $\Pi_{12}^{\mu\nu}$, which can be rewritten as
\begin{eqnarray}
\Pi_{12}^{\mu\nu} &=& \int_0^1dx\int_0^{1-x}dy\int_0^{1-x-y}dz\;\mu^{4-D}\int\frac{d^D\bar p}{(2\pi)^D}\nonumber\\
&&\times\frac{6ie^2}{(\bar p^2-M^2)^4}\mathrm{tr}\{[\bar q^2+ b^2+2(\bar q\cdot b)\gamma_5](\bar{\slashed{q}}+\slashed{b}\gamma_5) \nonumber\\
&&\times\bar\gamma^\mu[\hat{\slashed{p}},\slashed{b}]\gamma_5(\bar{\slashed{q}}_1+\slashed{b}\gamma_5)\bar\gamma^\nu\},
\end{eqnarray}
where we have considered $\hat{\slashed{q}}_1=\hat{\slashed{p}}$ (i.e., $\hat{\slashed{t}}_1=0$). Thus, its odd part, after the calculation of the trace, takes the form
\begin{eqnarray}\label{Pi12odd}
\Pi_{12|odd}^{\mu\nu} &=& -48e^2\int_0^1dx\int_0^{1-x}dy\int_0^{1-x-y}dz \\
&&\times\mu^{4-D}\int \frac{d^D\bar p}{(2\pi)^D}\frac{\epsilon^{\kappa\lambda\mu\nu} b_\kappa k_\lambda}{(\bar p^2-M^2)^4}(b^2+\bar q^2)\hat p^2. \nonumber
\end{eqnarray}
Now, when the integration over the momentum $\bar p$ is calculated and then the expansion around $D=4$ is taken into account, $\Pi_{b12|odd}^{\mu\nu}\to\Pi_{b12|CS}^{\mu\nu} $, with
\begin{eqnarray}\label{Pi12}
\Pi_{12|CS}^{\mu\nu} &=& \frac{ie^2}{4\pi^2}\epsilon^{\kappa\lambda\mu\nu}b_\kappa k_\lambda \int_0^1dx\int_0^{1-x}dy\int_0^{1-x-y}dz \hspace{0.7cm}\\
&&\times\bar g^{\alpha\beta}\hat g^{\alpha\beta}\left[-\frac{6}{\epsilon}+3\ln\left(\frac{M^2}{\mu'^2}\right)+\frac{b^2+t^2}{M^2}+1\right], \nonumber
\end{eqnarray}
where $\epsilon=4-D$ and $\mu'^2=4\pi\mu^2e^{-\gamma-i\pi}$. Finally, by considering the contraction $\bar g^{\alpha\beta}\hat g_{\alpha\beta}=D-4$ (i.e., $\bar g^{\alpha\beta}\hat g_{\alpha\beta}=-\epsilon$) and calculating the Feynman parameter integrals, from the first term of~(\ref{Pi12}), we obtain
\begin{equation}\label{Pi12CS}
\Pi_{12|CS}^{\mu\nu} = \frac{ie^2}{4\pi^2}\epsilon^{\kappa\lambda\mu\nu} b_\kappa k_\lambda,
\end{equation}
in the limit of~$D\to4$. Note that the other terms vanish in this limit.

Let us now consider the last contribution $\Pi_{22}^{\mu\nu}$, which, by employing the Feynman parameters, is written as
\begin{eqnarray}
\Pi_{22}^{\mu\nu} &=& \int_0^1dx\int_0^{1-x}dy\int_0^{1-x-y}dz\;\mu^{4-D}\int \frac{d^D\bar p}{(2\pi)^D} \\
&&\times\frac{6ie^2\mathrm{tr}\{[\hat{\slashed{p}},\slashed{b}](\bar{\slashed{q}}+\slashed{b}\gamma_5)\bar\gamma^\mu[\hat{\slashed{p}},\slashed{b}]\gamma_5(\bar{\slashed{q}}_1+\slashed{b}\gamma_5)\bar\gamma^\nu\}.}{(\bar p^2-M^2)^4}. \nonumber
\end{eqnarray}
After the calculation of the trace, only its odd part survives, which is given by
\begin{eqnarray}\label{Pi22odd}
\Pi_{22|odd}^{\mu\nu} &=& -96e^2\int_0^1dx\int_0^{1-x}dy\int_0^{1-x-y}dz \\
&&\times\mu^{4-D}\int \frac{d^D\bar p}{(2\pi)^D}\frac{\epsilon^{\kappa\lambda\mu\nu} b_\kappa k_\lambda}{(\bar p^2-M^2)^4}b^2\hat p^2, \nonumber
\end{eqnarray}
so that, in the following, after the calculation of the momentum integral, $\Pi_{22|odd}^{\mu\nu} \to \Pi_{22|CS}^{\mu\nu}$, where
\begin{eqnarray}\label{Pi22}
\Pi_{22|CS}^{\mu\nu} &=& \frac{ie^2}{2\pi^2}\epsilon^{\kappa\lambda\mu\nu}b_\kappa k_\lambda \int_0^1dx\int_0^{1-x}dy\int_0^{1-x-y}dz \nonumber\\
&&\times\bar g^{\alpha\beta}\hat g^{\alpha\beta} \frac{b^2}{M^2}.
\end{eqnarray}
Note that, as $\bar g^{\alpha\beta}\hat g^{\alpha\beta}=D-4$, in the limit of $D\to4$, trivially we have
\begin{equation}\label{Pi22CS}
\Pi_{22|CS}^{\mu\nu}=0.
\end{equation}
This is also what happens with the contributions coming from the second and other higher order terms of the propagator~(\ref{Bg5}), as discussed above.

Therefore, the final result is given by
\begin{eqnarray}\label{PiCS}
\Pi_{CS}^{\mu\nu}&=&\Pi_{11|CS}^{\mu\nu}+\Pi_{12|CS}^{\mu\nu}+\Pi_{21|CS}^{\mu\nu}+\Pi_{22|CS}^{\mu\nu} \nonumber\\
&=&\frac{ie^2}{2\pi^2}\epsilon^{\kappa\lambda\mu\nu} b_\kappa k_\lambda,
\end{eqnarray}
in which, by considering $S_{eff}^{(2)}\to S_{CS}$ in Eq.~(\ref{Seff2}), the Chern-Simons action becomes
\begin{equation}
S_{CS} = \frac{1}{2}\int d^4x\, \Pi_{CS}^{\mu\nu}(k\to i\partial) A_\mu A_\nu,
\end{equation}
with $(k_{AF})_\mu=-\frac{e^2}{4\pi^2}\,b_\mu$. Thus, as we have anticipated, the result $C_2=-\frac{e^2}{4\pi^2}$ is readily obtained when we use the `t Hooft-Veltman regularization, in particular, when we take into account  the new definitions of the anticommutation and commutation relations (\ref{gamma}) of the Dirac matrices.

\section{Finite temperature effect}\label{FT}

Our aim now is to discuss the finite-temperature effects on the coefficient $C$, as a natural extension of the present work. In order to implement this study, we first change the Minkowski space to Euclidean space (through the Wick rotation) and separate the space and time components of the momentum $\bar p$. This can be done by performing the following procedure: $p_0\to ip_0$ (or $\bar g^{\mu\nu}\to-\bar\delta^{\mu\nu}$), i.e., $\bar p^2\to -\bar\delta^{\mu\nu}\bar p_\mu \bar p_\nu=-\bar p^2$, $\bar p\cdot b\to-\bar p\cdot b$, $b^2\to-b^2$, and so on, as well as 
\begin{equation}
\mu^{4-D}\int \frac{d^D\bar p}{(2\pi)^D} = \mu^{3-d}\int \frac{d^d\vec p}{(2\pi)^d} \,i\int \frac{dp_0}{2\pi}
\end{equation}
and $\bar p^\mu=\vec p^\mu + p_0 u^\mu$, where $\vec p^\mu=(0,\vec p)$, $u^\mu=(1,0,0,0)$, and $D=d+1$, with $\vec p^\mu$ being a quantity in $d$ space dimensions.

The next step, due to the symmetry of the $\vec p$ integral under spatial rotations, is to consider the following substitutions:
\begin{eqnarray}
\vec p^\alpha \vec p^\beta &\to& \frac{\vec p^2}{d} (\bar\delta^{\alpha\beta}-u^\alpha u^\beta), \\
\vec p^\alpha \vec p^\beta \vec p^\gamma \vec p^\delta &\to& \frac{\vec p^4}{d(d+2)} [(\bar\delta^{\alpha\beta}-u^\alpha u^\beta)(\bar\delta^{\gamma\delta}-u^\gamma u^\delta) \nonumber\\
&&+(\bar\delta^{\alpha\gamma}-u^\alpha u^\gamma)(\bar\delta^{\beta\delta}-u^\beta u^\delta) \nonumber\\
&&+(\bar\delta^{\alpha\delta}-u^\alpha u^\delta)(\bar\delta^{\beta\gamma}-u^\beta u^\gamma)].
\end{eqnarray}
With these decompositions, we obtain covariant and noncovariant contributions to the polarization tensor (\ref{Pi}). Obviously, the noncovariant contributions vanish, when the $\vec p$ and $p_0$ integrals are performed. However, when we assume that the system is in thermal equilibrium with a temperature $T=\beta^{-1}$, the noncovariant contributions can eventually be nonzero. Indeed, in our case, only the noncovariant contributions will have an explicit dependence on temperature. Our main goal here is to study the behavior of the coefficient $C$ in the limit of high temperature, $T\to\infty$.

For this purpose, let us use the Matsubara formalism, which consist in taking $p_0=(n+1/2)2\pi/\beta$ and changing $1/(2\pi)\int dp_0\to 1/\beta \sum$, where the summation is carried out by using the following expression \cite{Ford:1979ds}:
\begin{equation}\label{sum}
\sum_n\bigl[(n+b)^2+a^2\bigl]^{-\lambda} = \frac{\sqrt{\pi}\Gamma(\lambda-1/2)}{\Gamma(\lambda)(a^2)^{\lambda-1/2}}+4\sin(\pi\lambda)f_\lambda(a,b)
\end{equation}
with
\begin{equation}\label{f}
f_\lambda(a,b) = \int^{\infty}_{|a|}\frac{dz}{(z^2-a^2)^{\lambda}}Re\Biggl(\frac{1}{e^{2\pi(z+ib)}-1}\Biggl).
\end{equation}
The above solution is valid only for $\lambda<1$, aside from the poles at $\lambda=1/2,-1/2,-3/2,\cdots$. Nevertheless, this restriction can be circumvented when we use the following recurrence relation
\begin{eqnarray}\label{rc}
f_{\lambda}(a,b) &=& -\frac1{2a^2}\frac{2\lambda-3}{\lambda-1}f_{\lambda-1}(a,b) \\
&&- \frac1{4a^2}\frac1{(\lambda-2)(\lambda-1)}\frac{\partial^2}{\partial b^2}f_{\lambda-2}(a,b) \nonumber
\end{eqnarray}
once, twice, and so on, until $\lambda$ is placed in the range of validity. 

By using the above considerations in the first contribution $\Pi_{11|odd}^{\mu\nu}$, Eq.~(\ref{Pi11odd}), and also taking into account for simplicity $k^2=0=b\cdot k$, as well as $k_0=0=b_0$, $\Pi_{11|odd}^{\mu\nu}\to\Pi_{11|CS}^{\mu\nu}$, where
\begin{eqnarray}\label{Pi11CST}
\Pi_{11|CS}^{\mu\nu} &=& 24e^2\int_0^1dx\int_0^{1-x}dy\int_0^{1-x-y}dz \nonumber\\
&&\times(\epsilon^{\kappa\lambda\mu\nu}b_\kappa k_\lambda \tilde I_1 + \epsilon^{0\lambda\mu\nu}b_0 k_\lambda \tilde I_2),
\end{eqnarray}
with
\begin{eqnarray}\label{tI1}
\tilde I_1 &=& \frac{i}{3}b^22^{2-d}\mu^{3-d}\pi^{-d/2}\Gamma\left(3-d/2\right)\frac{1}{\beta}\sum_n(p_0^2+M^2)^{\frac{d-8}{2}} \nonumber\\
&&\times\{-p_0^2[1-5z-5((1-x)x-2xz-z^2)] \nonumber\\
&&+b^2(1-x-z)(x+z)[2-4x-4z \nonumber\\
&&+4(x+z)^2-d(1-2x-2z)^2]\},
\end{eqnarray}
and
\begin{eqnarray}\label{tI2}
\tilde I_2 &=& \frac{i}{3}2^{-d}\mu^{3-d}\pi^{-d/2}\frac{1}{\beta}\sum_n(p_0^2+M^2)^{\frac{d-8}{2}} \nonumber\\
&&\{[6p_0^2-4b^2(1-2x-2z)^2]\Gamma\left(3-d/2\right)(p_0^2+M^2) \nonumber\\
&&+8b^2p_0^2(1-2x-2z)^2\,\Gamma\left(4-d/2\right) \nonumber\\
&&-3\,\Gamma\left(2-d/2\right)(p_0^2+M^2)^2\}.
\end{eqnarray}
In the above expressions we have performed the $\vec p$ integral. If we return to the $p_0$ integral, $1/\beta \sum\to1/(2\pi)\int dp_0$, after the calculation we obtain $\tilde I_1=0=\tilde I_2$, as expected.

Going back to Eqs.~(\ref{tI1}) and (\ref{tI2}), we can verify that, by calculating the Feynman parameters integrals, $\tilde I_1=0$, whereas $\tilde I_2\neq0$. Therefore, the temperature dependence only appears in the noncovariant contribution of Eq.~(\ref{Pi11CST}), in which by evaluating the summation of Eq.~(\ref{tI2}) (using Eq.~(\ref{sum})), we obtain
\begin{eqnarray}
\tilde I_2 &=& \frac{i}{24}\int_{|\xi|}^\infty d\zeta\frac{[\xi^2+2(1-x-z)(x+z)(3\zeta^2-5\xi^2)]}{(1-x-z)(x+z)(\zeta^2-\xi^2)^{1/2}} \nonumber\\
&&\times\tanh(\pi\zeta)\mathrm{sech}^2(\pi\zeta),
\end{eqnarray}
where $\xi = \beta M/2\pi$.

With regard to the other contributions, $\Pi_{12|odd}^{\mu\nu}$ and $\Pi_{22|odd}^{\mu\nu}$ (Eqs.~(\ref{Pi12odd}) and (\ref{Pi22odd}), respectively), we can easily observer that the noncovariant terms do not contribute because $\hat\delta^{\alpha\beta}u_\alpha=0$. Futhermore, the temperature-dependent terms of the remaining covariant contribution vanish, when we take into account $d=3$ in the contraction $\bar\delta^{\alpha\beta}\hat\delta_{\alpha\beta}=d-3$. Thus, only the temperature-independent and covariant terms survive, given the same results (\ref{Pi12CS}) and (\ref{Pi22CS}), respectively. 

The final result (\ref{PiCS}) is then rewritten as
\begin{equation}
\Pi_{CS}^{\mu\nu} = \frac{ie^2}{2\pi^2}\epsilon^{\kappa\lambda\mu\nu} b_\kappa k_\lambda + 24e^2\epsilon^{0\lambda\mu\nu} b_0 k_\lambda \tilde I_2.
\end{equation}
Thus, in the limit of high temperature, $\xi\to0$, after the evaluation of the integrals, 
\begin{eqnarray}\label{PiCST}
\Pi_{CS}^{\mu\nu}&\to&\frac{ie^2}{2\pi^2}\epsilon^{\kappa\lambda\mu\nu} b_\kappa k_\lambda - \frac{ie^2}{2\pi^2}\epsilon^{0\lambda\mu\nu} b_0 k_\lambda \nonumber\\
&=& \frac{ie^2}{2\pi^2}\epsilon^{i\lambda\mu\nu} b_i k_\lambda.
\end{eqnarray}
Therefore, at high temperature, $(k_{AF})_0\to0$ and $(k_{AF})_i\to-\frac{e^2}{4\pi^2}b_i$, meaning that the parity symmetry is restored in this regime. In the context of Weyl semimetals, these results are in accordance with the fact that the chiral magnetic current $j^\alpha=(k_{AF})_0\epsilon^{0\alpha\lambda\rho}\partial_\lambda A_\rho$ vanishes at high temperature \cite{Zyuzin:2012vn}, whereas the anomalous Hall current $j^\alpha=(k_{AF})_i\epsilon^{i\alpha\lambda\rho}\partial_\lambda A_\rho$ remains unaffected by the finite  temperature \cite{Goswami:2012db}. 

\section{Summary}

In this work, we have studied the radiative generation of the CPT-odd Chern-Simons term, the second term of Eq.~(\ref{lvqed}), arising from massless fermions (see Lagrangian~(\ref{fLp}) and discussion below). We have calculated the vacuum polarization tensor (\ref{Pi}) using 't Hooft-Veltman regularization scheme, in which we have obtained the result $C_2=-\frac{e^2}{4\pi^2}$ for the coefficient of the Chern-Simons term (see Eqs.~(\ref{PiCS})). This result leads us precisely to the same conductivity (\ref{cond}) found in Weyl semimetals, i.e., the 't Hooft-Veltman regularization scheme is the correct one to be used in this context. As a natural extension of the present work, we have also calculated the polarization tensor (\ref{Pi}) at finite temperature, in which at high temperature, $(k_{AF})_0\to0$ and $(k_{AF})_i\to-\frac{e^2}{4\pi^2}b_i$. In the context of Weyl semimetals, this indicates that the chiral magnetic current $\vec j=(k_{AF})_0\,\vec B$ vanishes at high temperature, whereas the anomalous Hall current $\vec j=(\vec k_{AF})\times\vec E$ remains unaffected by the temperature.

\begin{acknowledgments}
This work was supported by Conselho Nacional de Desenvolvimento Cient\'\i fico e Tecnol\'ogico (CNPq).
\end{acknowledgments}

\end{document}